\documentclass{mem}
\usepackage{txfonts}\usepackage{balance}
\usepackage{graphicx}
\usepackage[a4paper,breaklinks,dvipdfm]{hyperref}
\idline{0}{0}
\begin{document}
\def\teff{$T\rm_{eff }$}
\def\kms{$\mathrm {km s}^{-1}$}

\title{
UVES and FORS2 spectroscopy of the GRB081008 afterglow}

   \subtitle{}

\author{
V. \,D'Elia\inst{1,2}
\and S. \,Campana\inst{3} 
\and S. \,Covino\inst{3}
\and P. \,D'Avanzo\inst{3}
\and S. \,Piranomonte\inst{1}
\and G. \,Tagliaferri\inst{3}
          }

  \offprints{V. D'Elia}

\institute{
Istituto Nazionale di Astrofisica --
Osservatorio Astronomico di Roma, Via di Frascati 33,
I-00040 Monte Porzio Catone (RM), Italy
\and
ASI Science Data Center --
Via Galileo Galilei, I-00044 Frascati (RM), Italy
\and 
Istituto Nazionale di Astrofisica --
Osservatorio Astronomico di Brera, Via E. Bianchi 46,
I-23807 Merate (LC), Italy
\email{delia@asdc.asi.it}
}

\authorrunning{V. D'Elia }

\titlerunning{UVES and FORS2 spectroscopy of the GRB\,081008 afterglow}

\abstract{

  We study the GRB\,081008 environment with simultaneous high- and
  low-resolution spectroscopy using UVES and FORS2 data acquired $\sim
  5$ hr after the {\it Swift} trigger.  The interstellar medium (ISM)
  of the host galaxy at $z=1.9683$ is constituted by at least three
  components which contribute to the line profiles. Component I is the
  redmost one, and is $20$ km/s and $78$ km/s redward component II and
  III, respectively. We detect several ground state and excited
  absorption features in components I and II. These features have been
  used to compute the distances between the GRB and the
  absorbers. Component I is found to be $52 \pm 6$ pc away from the
  GRB, while component II presents few excited transitions and its
  distance is $200^{+60}_{-80}$ pc. Component III only features a few,
  low ionization and saturated lines suggesting that it is even
  farther from the GRB. The hydrogen column density associated to
  GRB\,081008 is $\log N_{\rm H}/{\rm cm}^{-2} = 21.11 \pm 0.10$, and
  the metallicity of the host galaxy is in the range [X/H] $= -1.29$
  to $-0.52$. In particular, we found [Fe/H]$=-1.19 \pm 0.11$ and
  [Zn/H]$=-0.52\pm 0.11$ with respect to solar values. This
  discrepancy can be explained by the presence of dust in the GRB ISM,
  given the opposite refractory properties of iron and zinc. By
  deriving the depletion pattern for GRB\,081008, we find the optical
  extinction in the visual band to be $A_V \sim 0.19$ mag. The Curve
  of Growth analysis applied to the FORS2 spectra brings column
  densities consistent at the $3\sigma$ level to that evaluated from
  the UVES data using the line fitting procedure. This reflects the
  low saturation of the detected GRB\,081008 absorption features.

}
\maketitle{}

\section{Introduction}

For a few hours after their onset, Gamma Ray Bursts (GRBs) are the
brightest beacons in the far Universe, offering a superb opportunity
to investigate both GRB physics and high redshift galaxies.  Early
time spectroscopy of GRB afterglows can give us precious information
on the kinematics, geometry, ionization and metallicity of the
interstellar matter of GRB host galaxies up to a redshift $z\sim5$, and
of intervening absorbers along the line of sight.  Our dataset
comprises nearly simultaneous UVES high resolution and FORS2 low
resolution spectra of GRB081008. High resolution spectroscopy is
important for many reasons: (i) absorption lines can be separated into
several components belonging to the same system; (ii) the metal column
densities can be measured through a fit to the line profile for each
component; (iii) fine structure and other excited lines can be
resolved. A comparison between high and low resolution data is
important to define a range of line parameters (column density as a
function of the oscillator strength and Doppler parameter) for which
the saturation effect can be correctly accounted for when high
resolution data are not available.

\begin{table}
\caption{Abundances computed from the UVES data}
{\footnotesize
\smallskip
\begin{tabular}{|l|cc|}
%\hline
\hline
Element $X$& $\log N_X /{\rm cm}^{-2}$    & $[X/{\rm H}]$    \\
\hline
O      & $>15.12\pm0.06$  & $>-2.68\pm0.11$ \\
Al     & $>13.70\pm0.04$  & $>-1.86\pm0.11$ \\
Si     & $ 15.75\pm0.04$  & $ -0.87\pm0.10$ \\
Cr     & $ 13.83\pm0.03$  & $ -0.92\pm0.10$ \\
Fe     & $ 15.42\pm0.04$  & $ -1.19\pm0.11$ \\
Ni     & $ 13.74\pm0.07$  & $ -1.29\pm0.12$ \\
Zn     & $ 13.15\pm0.04$  & $ -0.52\pm0.11$ \\
\hline
\end{tabular}
}
\end{table}

\begin{figure}[t!]
  \resizebox{\hsize}{!}{\includegraphics[clip=true]{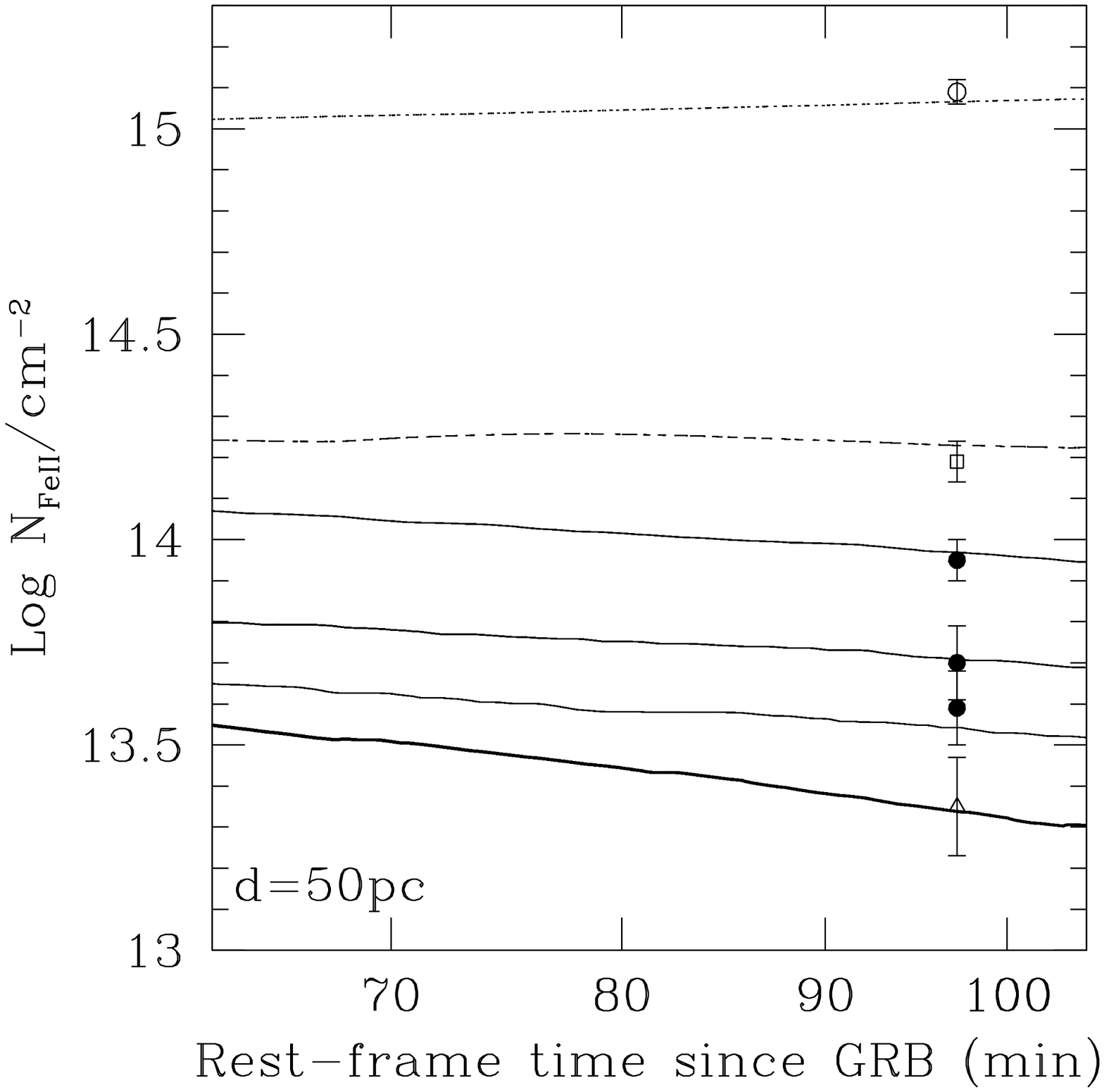}}
  \resizebox{\hsize}{!}{\includegraphics[clip=true]{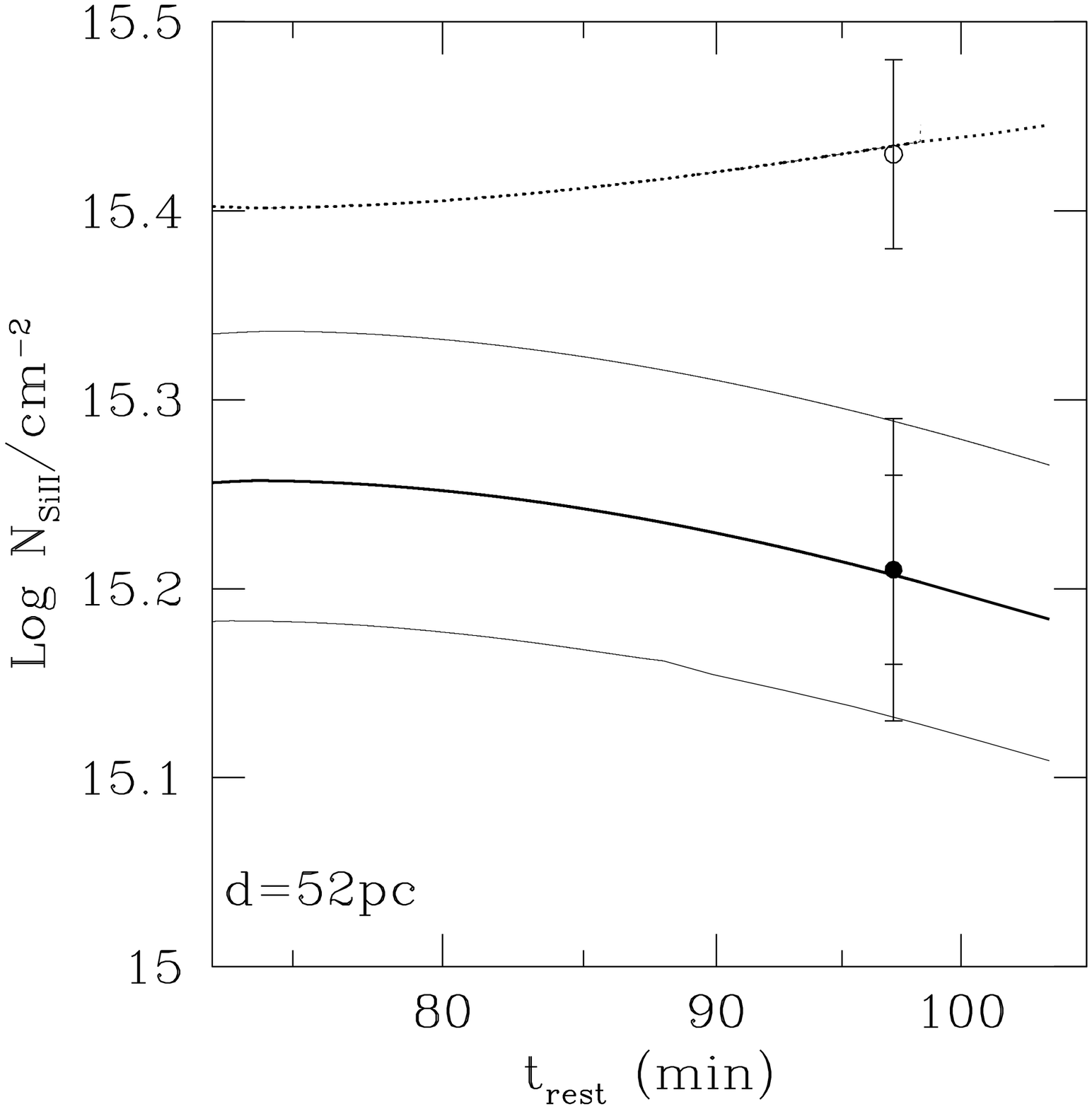}}
  \caption{ \footnotesize Comparison between the observed column
    densities in the FeII and SiII levels for component I of GRB
    081008 and that predicted by our photo-excitation code. A distance
    $of 51^{+21}_{-11}$ pc and $52\pm 6$ pc from the GRB is predicted
    using FeII (top panel) and SiII (bottom panel), respectively.  }
\label{eta}
\end{figure}

\section{Observations and Analysis}
The GRB081008 afterglow was observed with UVES $\sim4.30$ hours after
the trigger using dichroic $1$, for a total exposure of $1800$s. The
wavelength coverage is $3300-3900$ and $4800-6800$\AA, the spectral
resolution is $40.000$ and the S/N is $\sim 5$ and $\sim 8$ in the
blue and red band, respectively. Three FORS2 spectra of $900$s each,
acquired starting $4.37$ hours after the burst complete our
dataset. The wavelength coverage is $3500-6300$\AA, the spectral
resolution is $780$, and the S/N of the combined FORS2 spectrum is ~
$60-80$. The derived redshift of the host galaxy is $z=1.9683$. Just
two faint intervening absorbers are identified along the GRB 081008
sightline: one featuring the MgII 2796/2803 doublet at $z=1.286$ and
one featuring the CIV 1548/1550 doublet at $z=1.78$. UVES data were
analyzed using FITLYMAN in the MIDAS environment, to compute column
densities through Voigt fitting profile. Two components are necessary
in order to obtain a satisfactory fit for most of the line profiles of
the absorber at the redshift of GRB 081008. Actually, some strongly
saturated features show absorption in a third, bluemost component.

\section{Results}

\subsection{Metallicity}

Fitting the FORS2 Ly$\alpha$ feature we measure the HI column density:
log(N$_{HI}$/cm$^{-2}$)$=21.11\pm0.10$. Since the Ly absorption can not be
resolved into components (as for metals), we computed just an overall
metallicity, which is in the range $0.05-0.3$ with respect to solar
values (see table 1)

\begin{figure}[t!]
  \resizebox{\hsize}{!}{\includegraphics[clip=true]{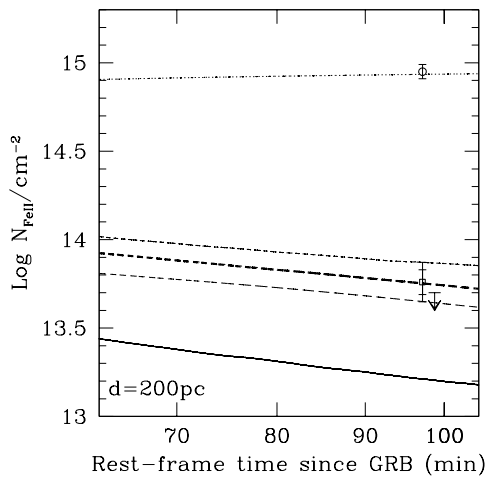}}
  \caption{ \footnotesize Comparison between the observed column
    densities in the FeII levels for component II of GRB 081008 and
    that predicted by our photo-excitation code. A distance of
    $200^{+60}_{-80}$ pc is predicted.  }
\label{eta}
\end{figure}

\subsection{Excited features and GRB/absorbers distance}

Excited features, belonging to the FeII, NiII and SiII levels have
been identified in GRB 081008. Vreeswijk et al (2007) and D’Elia et
al. (2009), using multi-epoch high resolution spectroscopy, observed
variability in the excited FeII and NiII lines of GRB 060418 and
080319B, which is a clear signature of indirect UV pumping exciting
such features. In other words, the decreasing UV flux coming from the
GRB excites the higher atomic levels with lesser and lesser
efficiency. The ratios between excited and ground state column
densities exclude collisional processes as responsible for the
production of excited features in GRB 081008, even if multi-epoch
spectroscopy to search for variability is not available. In this
scenario, the redmost component I is the closest to the GRB, since it
shows high absorption from the excited states. In order to compute
this distance we compare the observed columns with the results from a
time dependent photo-excitation code. We built a code that computes
the column densities of more than a hundred FeII and SiII levels as a
function of an incoming UV flux decreasing with time. Once the
lightcurve of GRB 081008 has been used as input for this code, we can
estimate the GRB/absorber distance. We find that the gas of component
I is $52\pm6$ pc away from the source (Fig. 1), while that of
component II is at $200^{+60}_{-80}$ pc (Fig.2). All errors are given
at the 90\% confidence level.

\subsection{FORS2 spectroscopy}

It is extremely interesting to compare low and high resolution
spectroscopic data in order to study the saturation problem. A line
that is saturated in a high resolution spectrum, may not appear
saturated in low resolution because its absorption is diluted in a
higher wavelength range. To compare UVES and FORS2 data, we used the
Curve of Growth analysis (COG, see Spitzer 1978) to compute the column
densities and b parameter from the equivalent widths (W$_r$) of the
FORS2 data (Fig. 3). Since FORS2 can not separate the absorption into
components, we summed up all the UVES contribution coming from
components I and II before comparing them with the FORS2 ones.  The
two sets of measurements are consistent at the $3\sigma$ level in the
worst case. This is not surprising, since UVES data are mostly not
saturated. However, this analysis provides a good set of column
density ranges for which the COG method applied to low resolution data
can provide reliable results.
 
\subsection{Dust depletion pattern}

Table 1 shows that FeII and ZnII abundances are significantly
different. This can be ascribed to the different refractory properties
of the two elements, with the former that preferentially tends to
produce dust grains while the latter prefers the gas phase. The
comparison between these opposite elements can thus provide
information on the dust content in the GRB environments. In order to
be more quantitative, we derive the dust depletion pattern for the GRB
081008 environment, following the method described in Savaglio
(2000). We consider the four depletion patterns observed in the Milky
Way, namely, those in the warm halo (WH), warm disk + halo (WHD), warm
disk (WD) and cool disk (CD) clouds. We find that the best fit to our
data is given by the WH cloud pattern, with a metallicity of
$Z_{GRB}/Z_{\sun} \sim 0.3$ (Fig. 4). This metallicity value is
consistent with our [Zn/H] measurement (Table 1). Since the latter
quantity is linked to the extinction (see e.g., Savglio, Fall \& Fiore
2003) we derive $A_V \sim 0.19$ mag along the GRB 081008 line of sight.

\begin{figure}[]
\resizebox{\hsize}{!}{\includegraphics[clip=true]{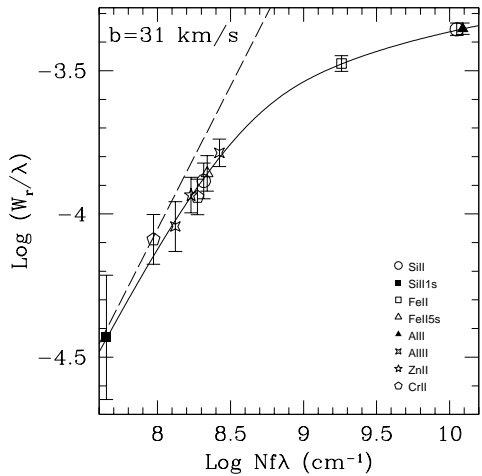}}
\caption{
\footnotesize
COG analysis applied to the FORS2 features with measured W$_r$. 
}
\label{li_vhel}
\end{figure}

\begin{figure}[]
\resizebox{\hsize}{!}{\includegraphics[clip=true]{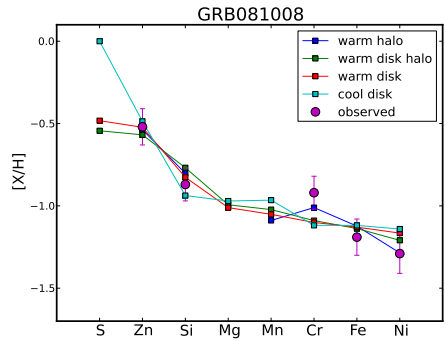}}
\caption{
\footnotesize
Depletion patterns in the GRB 081008 absorbing gas.
}
\label{li_vhel}
\end{figure}

\section{Conclusions}

UVES spectroscopy of GRB 081008 reveals a clumpy host galaxy gas with
at least three components. The host metallicity is in the range
$0.05-0.3$ (with respect to solar).

Excited lines detection and analysis set the GRB/absorber distance to
$\sim 50$ pc and $\sim 200$ pc for component I and II, respectively.

FORS2 spectroscopy complements high resolution data and provides a set
of column densities for which the COG method applied to low resolution
data is reliable.

The study of the dust depletion patterns enables to determine the
metallicity including dust. We estimate $Z_{GRB}/Z_{\sun}\sim0.3$ and
$A_V\sim0.19$.

More details can be found in D'Elia et al. 2011, MNRAS, in press).

\bibliographystyle{aa}

\end{document}